\input epsf
\input harvmac
\noblackbox


\def\narrowplus{\kern -.04truein + \kern -.03truein}
\def\narrowminus{- \kern -.04truein}
\def\narrowminussub{\kern -.02truein - \kern -.01truein}

\def\half{{1\over 2}}

\def\frac#1#2{{#1\over #2}}

%
%
\def\eqnn#1{\xdef #1{(\secsym\the\meqno)}\writedef{#1\leftbracket#1}%
\global\advance\meqno by1\wrlabeL#1}
\def\eqna#1{\xdef #1##1{\hbox{$(\secsym\the\meqno##1)$}}
\writedef{#1\numbersign1\leftbracket#1{\numbersign1}}%
\global\advance\meqno by1\wrlabeL{#1$\{\}$}}
\def\eqn#1#2{\xdef #1{(\secsym\the\meqno)}\writedef{#1\leftbracket#1}%
\global\advance\meqno by1$$#2\eqno#1\eqlabeL#1$$}

\lref\jp{J. Polchinski, D-branes and Ramond-Ramond charges,
hep-th/9510017, Phys. Rev. Lett. {\bf 75} (1995) 47.}
\lref\sag{A. Sagnotti, Open strings and their symmetry groups, in:
Non-perturbative Quantum Field Theory, Cargese 1987, eds. G.Mack
et. al. (Pergamon Press 1988)}
\lref\horava{P. Horava, Background duality of open string models,
Phys. Lett. {\bf B231} (1989) 251}
\lref\dlp{J. Dai, R.G. Leigh and J. Polchinski, New connections 
between string theories, Mod. Phys. Lett., {\bf A4} (1989) 2073}
\lref\polch{J. Polchinski, S. Chaudhuri and C.V. Johnson, Notes on
D-branes, hep-th/9602052}
\lref\witab{E. Witten, On S-duality in abelian gauge theory, 
hep-th/9505186.} 
\lref\witt{ E. Witten, Five branes and M-theory on an orbifold,
hep-th/9512219, Nucl. Phys. {\bf B463} (1996) 383.}
\lref\witten{E. Witten, Flux quantization in M-theory and the 
effective action, hep-th/9609122.}
\lref\hw{P. Horava and E. Witten, Heterotic and type I string dynamics
from eleven dimensions, hep-th/9510209, Nucl. Phys. {\bf B460} (1996) 
506; Eleven-dimensional Supergravity on a manifold with boundary, 
hep-th/9603142.}
\lref\kdsmorb{K. Dasgupta and S. Mukhi, Orbifolds of M-theory,
hep-th/9512196, Nucl. Phys. {\bf B465} (1996) 399}
\lref\kdsm{K. Dasgupta and S. Mukhi, F-theory at constant coupling, 
hep-th/9606044, Phys. Lett. {\bf B385} (1996) 125.}
\lref\casn{C. Ahn and S. Nam, Compactifications of F-theory on Calabi-Yau 
three-folds at constant coupling, hep-th/9701129.}
\lref\dj{D. P. Jatkar, Non-perturbative enhanced gauge symmetries in the
Gimon-Polchinski Orientifold, hep-th/9702031.}
\lref\kdsmI{K. Dasgupta and S. Mukhi, A note on low-dimensional string 
compactifications, hep-th/9612188, Phys. Lett. {\bf B398} (1997) 285.}
\lref\senD{A. Sen, Duality and orbifolds, hep-th/9604070, Nucl. Phys. 
{\bf B474} (1996) 361.}
\lref\sen{ A. Sen, F-theory and orientifolds, hep-th/9605150, Nucl.
Phys. {\bf B475} (1996) 562.}
\lref\dmns{D. Morrison and N. Seiberg, Extremal transitions and 
five-dimensional supersymmetric field theories, hep-th/9609070.}
\lref\ghm{M.B. Green, J.A. Harvey and G. Moore, I-brane inflow and 
anomalous couplings on D-branes, hep-th/9605033, 
Class. Quant. Grav. {\bf 14}(1997) 47.}
\lref\hm{J.A. Harvey and G. Moore, Five-brane instanton and $R^2$ 
couplings in N=4 string theory, hep-th/9610237.}
\lref\hmI{J.A. Harvey and G. Moore, Exact gravitational threshold 
correction in the FHSV model, hep-th/9611176.}
\lref\svw{S. Sethi, C. Vafa and E. Witten, Constraints on low-dimensional 
string compactifications, hep-th/9606122.}
\lref\bsv{M. Bershadsky, V. Sadov and C. Vafa, D-branes and
topological field theories, hep-th/9611222, Nucl. Phys. 
{\bf B463} (1996) 420.}
\lref\bottu{R. Bott and L. Tu, Differential forms in algebraic topology,
Springer-Verlag.}
\lref\vawitt{C. Vafa and E. Witten, A strong coupling test of string
duality, hep-th/9408074, Nucl. Phys. {\bf B431} (1994) 3.}
\lref\seitwo{N. Seiberg, IR dynamics on branes and space-time geometry,
hep-th/9606017, Phys.Lett. {\bf B384} (1996) 81.} 
\lref\seifour{N. Seiberg, Five-dimensional SUSY field theories, 
nontrivial fixed points and string dynamics, hep-th/9608111,
Phys.Lett. {\bf B388} (1996) 753.}
\lref\seiwitthree{N. Seiberg and E. Witten, Gauge dynamics and
compactification to three dimensions, hep-th/9607163.}
\lref\mill{J. W. Milnor and J. D. Stasheff, Characteristic Classes,
section 4, Princeton University Press (1974) 37-50.}
\lref\bfss{T. Banks, W. Fischler, S.H. Shenker and L. Susskind,
M-theory as a matrix model: a conjecture, hep-th/9610043,
Phys. Rev. {\bf D55} (1997) 5112.}
\lref\seibtf{N. Seiberg, New theories in six dimensions and matrix 
description of M-theory on $T^5$ and $T^5/Z_2$, hep-th/9705221.}
\lref\morales{J. F. Morales, C. A. Scrucca and M. Serone, Anomalous 
Couplings for D-branes and O-planes, hep-th/9812071.}
\lref\stefanski{B. Stefanski Jr., Gravitational 
Couplings of D-branes and O-planes, hep-th/9812088.}
\lref\craps{B. Craps and F. Roose, 
(Non)-Anomalous D-brane and O-plane Couplings: The normal bundle, 
hep-th/9812149.}

\Title{\vbox{\hbox{hep--th/9707224}\hbox{TIFR/TH/97-41}
\hbox{MRI-PHY/P970718}\hbox{UVA-ITFA/97/29}}}
{Gravitational Couplings and $Z_2$ Orientifolds}

\centerline{Keshav Dasgupta${}^1$, Dileep P. Jatkar${}^2$ and 
Sunil Mukhi${}^{3,1}$\footnote{$^*$}{e-mail: keshav@theory.tifr.res.in, 
dileep@mri.ernet.in, mukhi@theory.tifr.res.in}}
\medskip
\centerline{\it ${}^1$Theoretical Physics Group}
\centerline{\it Tata Institute of Fundamental Research}
\centerline{\it Homi Bhabha Road, Mumbai 400 005, INDIA}
\medskip
\centerline{\it ${}^2$Mehta Research Institute of Mathematics} 
\centerline{\it and Mathematical Physics}
\centerline{\it Chhatnag Road, Jhusi, Allahabad 221 506, INDIA}
\medskip
\centerline{\it ${}^3$Instituut voor Theoretische Fysica}
\centerline{\it Universiteit van Amsterdam}
\centerline{\it Valckenierstraat 65, 1018 XE Amsterdam, NETHERLANDS }

\vskip .2in
The interplay between gravitational couplings on branes and the
occurrence of fractional flux in low dimensional orientifolds is
examined. It is argued that gravitational couplings need to be
assigned not only to D-branes but also to orientifold planes. The
fractional charges of the orientifold $d$-planes can be understood in
terms of flux quantization of the $d-3$ form potential and modified
Bianchi identities. Detailed results are presented for the case of the
type IIB orientifold on $T^6/Z_2$, which is dual to F-theory on a
complex 4-fold with terminal singularities.

\Date{7/97}

\newsec{Introduction}

Recent developments have demonstrated the importance of a variety of
extended dynamical objects, branes\refs{\jp}, in string theory. One
context where (Dirichlet) branes arise naturally is in the
construction of orientifolds\refs{\sag,\horava,\dlp}, where one can
sometimes think of them as twisted sector states. Another kind of
object that occurs in orientifold constructions is called the
orientifold plane, the locus of fixed points of some discrete group.

Planes are usually assumed to be non-dynamical, as indeed they are at
weak coupling. But it has been shown in a few contexts\refs{\sen,
\dmns} that at strong coupling they can behave like
dynamical objects. Another context where the distinction between
branes and planes is blurred is in F-theory compactifications at
constant self-dual coupling\refs{\kdsm,\casn,\dj}. Here, F-theory
branes move around in groups and can even produce exceptional
symmetries, yet these configurations are continuously connected within
the constant-coupling moduli space to the perturbative configurations
of branes and planes.

An essential distinction between branes and planes is that the former
carry Yang-Mills gauge fields on their world-volume, and have moduli
for their locations, while the latter do not. Here we want to focus on
a complementary feature in which some amount of symmetry is maintained
between the two types of objects. Besides Yang-Mills couplings, branes
also carry gravitational couplings localized on their world-volume.
As we will see, orientifold planes also carry such localized
gravitational couplings, essentially because they are loci of singular
curvature. This fact neatly fits in with the various observations
referred to above regarding the strong-coupling behaviour of planes.

An interplay between gauge and gravitational couplings is a key
feature in maintaining consistency and anomaly-freedom in string
compactifications. Hence exploring the apparently distinct origins of
the two in theories with branes and planes may tell us something quite
fundamental about the underlying theory.

Another, apparently unrelated, issue in orientifolds is that while
branes always carry integer charge with respect to some $p$-form gauge
field, planes can carry fractional charge in a very precise way. We
will see that the presence of gravitational couplings on planes and
branes together conspires to explain this fact and render these
fractional charges consistent with Dirac quantization.

The appearance of fractional charges on planes can be argued for many
low dimensional compactifications of M-Theory and string
theory\refs{\polch}. The most straightforward way to see this is by
examining the orientifold of the type II string on $T^n/Z_2$ where $n$
is even for IIB and odd for IIA. There are always 16 D-branes in the
vacuum and $2^n$ orientifold planes. Symmetry and charge conservation
dictate that each plane carries $2^{4-n}$ units of charge. This is
fractional as soon as $n$ is greater than 4.

Analogous situations occur in M-theory. In the $T^5/Z_2$ orientifold,
the fact that twisted-sector states are outnumbered by the fixed
planes was noted and studied in Ref.\refs\kdsmorb. A careful analysis
of this situation in Refs.\refs{\witt,\witten} revealed that as in
stringy orientifolds, the planes carry fractional charge, and
explained how this is in fact consistent.  

The origin of fractional charge in this case is as follows. The
$T^5/Z_2$ compactification has 32 fixed points. Anomaly cancellation
requires 16 copies of $N=2$ tensor multiplets in six
dimensions. Assuming each tensor multiplet comes from a space-filling
5-brane, this would give rise to two interesting phenomena: (i) the
$C^{(3)}\wedge I_8$ term from the $D=11$ supergravity ($C^{(3)}$ being
the three form of M-theory) cancels anomalies locally on the brane by
anomaly inflow, and (ii) a magnetic charge $+1$ appears on each of the
5-branes.  Therefore charge cancellation requires the planes to carry
$-{1\over 2}$ $C^{(3)}$-field magnetic charge while the anomalies
automatically cancel locally, on both branes and planes, by the inflow
term in the lagrangian.

Some more interesting cases arise in low-dimensional compactifications
of string theory and M-theory. For example, consider the type II
theories on $T^8/Z_2$ orbifolds and orientifolds. Type IIB theory on
the $T^8/Z_2$ orientifold has 256 orientifold planes and 16
D1-branes. Charge cancellation would require orientifold planes to
carry $-{1\over 16}$ units of $\tilde B_{\mu\nu}$ charge.

Type IIA on $T^8/Z_2$ orbifold has 256 twisted sectors, but they do
not contribute any massless multiplets. The massless states in the
twisted sector can only arise in the RR sector since the left or right
moving fermions in the NS sector give vacuum energy greater than
zero. But the RR ground state in this case does not survive GSO
projection\refs{\senD}.  However, due to the existence of a
$B_{\mu\nu}$ tadpole in two dimensions\refs{\svw} a consistent
compactification requires $\chi/24$ one-branes (fundemental type IIA
strings) to condense in the vacuum, where $\chi$ is the Euler
characteristic of the compact manifold. For $T^8/Z_2$ the orbifold
Euler characteristic is 384, and thus tadpole cancellation requires 16
type IIA strings in the vacuum. Then charge cancellation would require
the fixed points to carry $-{1\over 16}$ units of $B_{\mu\nu}$
charge. 

This lifts to M-theory and F-theory on the same orbifold. In
the M-theory case the planes have $-{1\over 16}$ units of 3-form
charge while in F-theory there are really only $2^6$ rather than $2^8$
fixed points (we count only fixed points on the base) and these carry
$-{1\over 4}$ units of 4-form charge. Indeed, this is in the class of
$T^n/Z_2$ orientifolds referred to above, with $n=6$. Note that the
orbifold $T^8/Z_2$ has terminal singularities, and hence requires
irrelevant rather than marginal operators in order to be blown
up. String and M propagation on it, however, appear to be smooth.

There is also a dual pair with chiral supersymmetry in
$1+1$ dimensions. Type IIB on the $T^8/Z_2$ orbifold is chiral and has
potential gravitational anomalies. The 256 twisted sectors carry a
total gravitational anomaly of ${64\over 3}$ (in the units of
Pontryagin numbers) and this is cancelled by the total anomaly of 256
chiral bosons from the fixed points, which is $-{64\over
3}$\refs{\kdsmI}. The same anomaly cancellation occurs for M-Theory on
$T^9/Z_2$ (which is conjectured to be dual to IIB on $T^8/Z_2$) with
some crucial differences. The fixed points, 512 in number, now give
chiral fermions and their anomalies cancel the anomalies coming from
the untwisted sectors.

Another low-dimensional case is type IIA on the $T^9/Z_2$
orientifold. This has the usual 16 D0-branes in the vacuum and 512
orientifold points.  Charge cancellation would now require the
orientifold points to carry $-{1\over 32}$ units of $A_{\mu}$ charge.
We will make some new observations about these low-dimensional cases
later on.

There is an interesting relationship between Bianchi identities and
flux quantization that we will exploit in this paper. The Bianchi
identity of a $p$-form potential (in type IIA or IIB strings or in
M-theory) in $d$ uncompactified dimensions is related to charge
quantization of the field in $d-4$ uncompactified dimensions. As an
example, the Bianchi identity of the $C^{(3)}$ field of M theory on
$S^1/Z_2$ is related to the flux quantization of $G^{(4)} = dC^{(3)}$
for M theory on $T^5/Z_2$\refs{\witten}.After compactification on a
circle this descends to an analogous relation for type IIA on
$S^1/Z_2$ and $T^5/Z_2$. Other examples include the flux quantization
of $G^{(5)} = dC^{(4)}$ for type IIB compactified on $T^6/Z_2$
orientifold ($C^{(4)}$ is the self-dual RR four-form potential of type
IIB). This is related to the Bianchi identity of type IIB on the
$T^2/Z_2$ orientifold. Similarly flux quantization for IIA on
$T^7/Z_2$, IIB on $T^8/Z_2$ and IIA on $T^9/Z_2$ is related to Bianchi
identities on $T^3/Z_2$, $T^4/Z_2$ and $T^5/Z_2$ respectively.

This paper is organized as follows. In Section 2 we argue that
orientifold planes carry gravitational WZ terms, and discuss how
Bianchi identities get modified in the presence of planes and
branes. In Section 3 we show that the modified Bianchi identities
indeed lead to consistent behaviour of branes when transported around
planes even if the latter carry fractional charge, so that Dirac
quantization is always satisfied. In Section 4 we discuss Wess-Zumino
terms of $R^4$ type that are supported on planes as well as branes.
In Section 5 we focus on the special case of the $T^6/Z_2$
orientifold, where 3-branes condense in the vacuum. Gravitational
couplings in 3+1 dimensional supersymmetric gauge theories have
received some attention recently\refs{\hm,\hmI,\witab}. We interpret
our results in the context of N=4 compactifications. Finally in
Section 6 we comment on low-dimensional cases.

\newsec{Orientifolds and Fractional Charges}

Orientifolds are constructed by gauging the world sheet parity
transformation along with some target space discrete symmetry of type
II string theory.  This gauging gives non-vanishing disc
tadpoles. That means orientifold planes are charged with respect to
the field whose disc tadpole is non-vanishing. These charges can be
cancelled by inserting an appropriate number of space-filling
D-branes. In the simplest case, the $Z_2$ orientifolds,
we need 16 D-branes to cancel the charge carried by orientifold
planes. 

These orientifolds can also be understood via T-duality. Consider type
IIB string theory in ten dimensions. Orientifolding this theory in ten
dimensions gives the type I string. All other $Z_2$ orientifolds in
lower dimensions can then be understood by T-dualizing type I string
theory after toroidal compactification. In ten dimensions, the
orientifold 9-plane carries charge $-16$ with respect to the 10-form
potential.  This is cancelled by condensing 16 D-9 branes in the
vacuum, which gives the well known SO(32) gauge group of type I string
theory. The orientifold 9-plane splits into two orientifold 8-planes
after compactification and T-duality on a circle. These orientifold
8-planes carry charge $-8$ each with respect to 9-form potential. Thus
again we need 16 D-8 branes to cancel the charge on the orientifold
planes.

A special vacuum is the one for which the D-8 branes cancel the
orientifold charge locally. In this case the 16 D-8 branes are placed
on the two orientifold planes in bunches of 8 each.  As we compactify
further and T-dualize along the compact directions, the number of
orientifold planes keep doubling with each action of T-duality,
whereas the total charge carried by all the orientifold planes remains
equal to $-16$ which is equally distributed among them. Thus for
compactifications on higher-dimensional tori, orientifold planes carry
fractional charge, and special vacua with local charge cancellation do
not exist unless the compactification tori are squashed to merge
orientifold planes. (It is intriguing that this occurs just at the
value of uncompactified dimension (6) below which the m(atrix) theory
proposal\refs{\bfss} starts to become problematic\refs{\seibtf}.)

The first instance where fractional charges on the orientifold planes
occur is the $T^5/Z_2$ orientifold compactification of type IIA string
theory. Using the relation of M-theory and type IIA string theory, the
same conclusion can be reached for the $T^5/Z_2$ orientifold of
M-theory.  In both cases, the orientifold planes carry half-integral
magnetic charge with respect to the three form field $C^{(3)}$. This
phenomenon was explained by Witten\refs{\witten}, who showed that the
fluxes of the four form field strength $G^{(4)}=dC^{(3)}$ are
quantized in half-integral units. Whenever the $G$ flux through a
four-cycle $M$ of an eleven-dimensional manifold $Y$ is half integral,
the first Pontryagin class $p_1(Y)$ of $Y$ restricted to $M$ is an
integer divisible by two but not by four. The flux which is integrally
quantized belongs to $G^{(4)}-(p_1(Y)/4)$.

Since orientifolds produce vacuum charges which are canceled by
branes, the space-time action for the orientifolds can be written as
\eqn\action{I_{orient} = I_{bulk}+ \sum_{i=1}^{16} I^{(i)}_{DBI},}
where the first term on the right hand side is the bulk space-time
action of type IIA(IIB) string theory, subject to the orientifold
projection, and the second term is the sum over Dirac-Born-Infeld
actions for the 16 space-filling D-branes coupled to type IIA(IIB) 
potentials. 

The first term, the bulk string theory action, is written in ten
dimensions, whereas the space-filling D-brane actions fill the space
transverse to $T^n/Z_2$. The sum is taken over 16 points on $T^n/Z_2$
where these D-branes are localised, hence in fact the second term on
the RHS of \action\ is accompanied by $n$ dimensional
$\delta$-functions specifying locations of D-branes on $T^n/Z_2$. We
will always consider curved non-compact space, i.e., both orientifold
planes and the D-brane world volumes are curved.

As mentioned above, in curved space the D-brane world-volume has
additional couplings. These are Wess-Zumino terms which are wedge
products of the $p$-form field with powers of the curvature two-form
$R$. In particular, on the worldvolume of a D-5-brane in curved space,
there is an additional Wess-Zumino term coupling the RR two-form $B$
to the first Pontryagin class $p_1(R)$\refs{\bsv}. One can see this by
a simple anomaly inflow argument\refs{\ghm}.

Now consider the orientifold of type IIB string theory in ten
dimensions. In this case we have an orientifold 9-plane accompanied by
16 D-9-branes, leading to type I string theory. The modified Bianchi
identity for the three-form field strength due to the anomaly
cancellation condition in the type I string is
\eqn\bianchi{dH={1\over 2}~[p_1(R) - p_1(F)].}
where $p_1(R)$ is the first Pontryagin class of the spin manifold Y
and is defined as ${1\over 8\pi^2}~ \tr (R\wedge R)$ and similarly
$p_1(F) = {1\over 8\pi^2}~\tr (F\wedge F)$. For a spin manifold, $p_1$
is divisible by 2, hence $\lambda \equiv {p_1\over 2}$ defines an
integer cohomology class.  Let us try to understand this equation from
the orientifold point of view.  The D9-brane worldvolume action is
given by
\eqn\dbi{S_{D9}= S_{DBI}+S_{WZ},}
where $S_{BI}$ is the usual Dirac-Born-Infeld action and $S_{WZ}$ is the
contribution of the Wess-Zumino terms. We will not write all these terms 
explicitly. The relevant Wess-Zumino terms\refs{\ghm}
are\footnote{$^1$} {By the anomaly inflow argument, the WZ
term that actually occurs on the brane world volume is proportional to
$$\int_{B_p} C\wedge \tr_n~ exp({F\over 2\pi}) \sqrt {{\hat A}(R)}$$
where ${\hat A}(R) = 1- {p_1\over 24} + {7p_1^2-4p_2\over 5760} + ...$
and $p_i$ are the Pontryagin numbers. For our case it suffices to take
$C = {^*B}$.}

\eqn\wz{\int {^*B}\wedge~{1\over 16\pi^2}~\tr(F\wedge F)\quad 
{\rm and}\quad 
\frac{1}{24}\int {^*B}\wedge~{1\over 16\pi^2}~\tr(R\wedge R).}
where $^*B$ is the Poincare dual of the RR two form $B$.

We also have the space-time action of type IIB string theory. The term 
relevant for our purpose is
\eqn\toB{S_{IIB}\sim {1\over 2}\int H^2+\cdots,}
where, $H$ is the field strength of the RR sector two-form field
$B$. The equation of motion for $^*B$ would then be given by
\eqn\eom{dH = {1\over 16\pi^2}(-\tr F\wedge F +\frac{2}{3} 
\tr R\wedge R)} 
where the RHS is the contribution coming only from 16
D-9-branes. Since orientifold planes are not dynamical they do not
couple to $F$ but since we are considering both D-branes as well as
orientifold planes in curved space, planes can couple to $R$. We
claim that orientifold planes contribute a further
\eqn\rr{\frac{1}{3}~{\tr(R\wedge R)\over 16\pi^2} = {p_1\over 6}}
to the RHS of the above equation.

One way to see this is the following. In case of D-branes both the
terms in \wz\ occur at the disc order with three insertions. For the
orientifold plane also these terms should contribute at the same
order, except that now the disc is replaced by $RP^2$. The first term
in Eq.\wz\ requires two open-string insertions, whereas the second
term has all three closed-string insertions. Since $RP^2$ has no
boundary and open string vertices are inserted on boundaries, $RP^2$
does not contribute to the first term, which is equivalent to the
statement that the orientifold planes have no open-string dynamics or
they do not couple to Yang-Mills fields. Closed-string vertex
insertions are in the bulk of the worldsheet and hence are allowed on
$RP^2$. Thus the three point vertex on $RP^2$, i.e., the orientifold
plane, contributes a term proportional to the second term in
Eq.\wz. Both the disc and $RP^2$ are tree-level diagrams, and at tree
level only D-branes and orientifold planes can contribute these
terms. Since we already know the modified Bianchi identity as well as
the contribution of D-branes, the term Eq.\rr\ has to come from the
orientifold plane. We will see that this interpretation makes sense
when we consider other cases, particularly the 8-dimensional example.

Though the number of orientifold planes multiplies on compactification
followed by T-duality, the total contribution of orientifold planes
towards the appropriate Bianchi identity remains the same. In other
words, if $C^{(n)}$ is an RR $n$-form, then the ${1\over 3}C^{(n)}\wedge
{tr(R\wedge R)\over 16\pi^2}$ term residing on the orientifold planes
is equally distributed among all the orientifold planes and the total
contribution of orientifold planes to the Bianchi identity for
$C^{(n)}$ is equal to Eq.\rr . In the case of D-5-branes,
Ref.\refs{\bsv} could not fix the sign of the Wess-Zumino term
containing the Pontryagin class.  Relating this term to the Bianchi
identity, it is easy to see that there is a relative minus sign
between the two terms occuring in \wz . Incorporating this
contribution of orientifold planes gives us the correct Bianchi
identity
\eqn\bid{dH = {1\over 16\pi^2}~ (-\tr F\wedge F + \tr R\wedge R).}

Another related way of seeing this is the following. In ten
dimensions, the 3-form field strength in the type-I string has a
kinetic term
\eqn\kinetic{S_I = \half \int H\wedge *H}
where
\eqn\hdef{
H = dB + \omega_{3L} - \omega_{3Y}.}
Here, $d\omega_{3L} \equiv {1\over 16\pi^2}~\tr R\wedge R$ and
similarly for $d\omega_{3Y}$ with R replaced by F. This leads to a
cross term
\eqn\cross{
\int *dB\wedge (\omega_{3L} - \omega_{3Y})}
Now let $B^{(6)}$ be the dual of $B$ defined by $*dB = dB^{(6)}$. Then
the above coupling becomes the WZ term
\eqn\dualcoup{
{1\over 16\pi^2}~\int B^{(6)}\wedge \left(\tr(F\wedge F) - 
\tr(R\wedge R)\right) }
where $B^{(6)}$ is a (dual) RR potential in type I.

{}From the coefficient of the term above, it is evident that the
curvature terms have the right coefficient to arise from 24
9-branes. But in reality there are only 16 9-branes, which contribute
2/3 of the desired factor, and the remaining 1/3 is ascribed to the
planes as in Eq.\rr.

Consider now compactification of type I theory on a circle followed by
T-duality. This gives us type I$'$ theory which can also be obtained
from type IIA theory on an $S^1/Z_2$ orientifold. As mentioned
earlier, T-duality doubles the number of orientifold planes. In this
case we get two orientifold planes and \rr\ is distributed equally
between them. The vacuum with local charge cancellation is the one
where eight D8 branes are located on one orientifold plane and eight
on the other. Let us focus only on one of the orientifold planes. The
total orientifold action (modulo projection) in the vicinity of one
orientifold plane is
\eqn\sfull{
S_{orient} = S_{IIA}+ S_{D8}.}
The relevant terms in this action are:
\eqn\onepr{
S_{orient} \sim {1\over 2}\int G^2 + \int  {^*C^{(5)}}
\wedge{\tr(F\wedge F)\over
16\pi^2}\delta(x^9)
-8\int \frac{1}{24}{^*C^{(5)}}\wedge {\tr(R\wedge R)\over 16\pi^2}
\delta(x^9),}
where $x^9$ is the compact circle coordinate, the orientifold plane
that we are concentrating on is localised at $x^9=0$, $^*C^{(5)}$ is
the 5-form dual to the RR three-form potential $C^{(3)}$ in type IIA
string theory. The field strength $G^{(4)}=dC^{(3)} + \ldots$ where
extra terms have to be introduced to ``solve'' the Bianchi identity as
we discuss below. The gauge field strength $F$ takes values in the
group SO(16) whereas the curvature ($R$) terms all add up, so that the
WZ term of a single D8-brane is multiplied by 8. Lastly the $\delta$
function tells us that the branes are orthogonal to $x^9$.

The equation of motion for the field $^*C^{(5)}$ is given by
\eqn\nubianki{
dG^{(4)} = {1\over 16\pi^2}~(\frac{1}{3}\tr R\wedge R - \tr F\wedge F)
\delta(x^9)}
where the RHS is a contribution coming entirely from the branes.  We
have argued that the $R\wedge R$ contribution from the orientifold
planes is equally distributed among the planes. In the present case, a
single orientifold plane contributes ${1\over 6}~{\tr(R\wedge R)\over
16\pi^2}\delta(x^9) = {p_1\over 12}\delta(x^9)$. Adding this
contribution to the Bianchi identity of $G$ we get
\eqn\horwit{
dG^{(4)} = {1\over 16\pi^2}~(\frac{1}{2}\tr R\wedge R-
\tr F\wedge F)\delta(x^9).}
This is the analogue, for type IIA string theory on $S^1/Z_2$, of an
equation derived in the second paper of Ref.\refs{\hw} in the context
of the strong coupling limit of this theory, namely M-theory on
$S^1/Z_2$. That equation was used in Ref.\refs{\witten} to show that
$G^{(4)}$ can have half integral fluxes in M-theory.

Despite the similarity in the final equation, there is an important
difference between the derivation of our result and that in
Ref.\refs{\hw}. In the latter case, there are really no branes, just
fixed planes, since there are no moduli to break $E_8\times E_8$. For
string theory orientifolds below 10 dimensions, perturbatively there
are always branes and planes, and they can be separated from each
other. Hence it is essential to understand the contribution of each
one separately, to obtain the correct Bianchi identity in the special
charge-cancelling configuration as we have just done. An essential
role was played here by the gravitational coupling on the fixed
planes.

Now we turn to the $T^2$ compactification of type I string
theory. This is equivalent, by T-duality, to the $T^2/Z_2$ orientifold
of type IIB strings. This model has been studied in great detail by
Sen\refs{\sen}. Here we have four orientifold planes which carry
seven-brane charge $-4$. This charge can be canceled by putting four
seven branes on the top of each orientifold plane. From what we have
said earlier every orientifold plane in this case will contribute a
factor ${1\over 12}~{\tr(R\wedge R)\over 16\pi^2}\delta(x^8)\delta(x^9)
= {p_1\over 24}\delta(x^8)\delta(x^9)$. This is consistent with
results of \refs{\sen}: when seven branes are taken away from the
orientifold plane, the plane splits into two seven branes which are
mutually non-local and also non-local with respect to the original
seven branes. The curvature term that we expect from the orientifold
plane is exactly twice as much as that contributed by a single
D-brane. We therefore see that each orientifold plane in this case can
split into two 7-branes which share the curvature terms. Once we take
this into account the Bianchi identity becomes
\eqn\bate{dG^{(5)} = {1\over 16\pi^2}~(\frac{1}{4}\tr R\wedge R - 
\tr F\wedge F)
\delta(x^8)\delta(x^9),}
where $G^{(5)}=dC^{(4)}+\ldots$ and $C^{(4)}$ is the self-dual RR
four-form in type IIB string theory. Although the 24 7-branes of
F-theory that emerge from the nonperturbative analysis carry different
$(p,q)$ charges, hence are related to each other by $SL(2,Z)$
S-duality, they all must carry the same worldvolume term contributing
to the above Bianchi identity since $C^{(4)}$ is
$SL(2,Z)$-invariant. This is a nice confirmation that in this situation,
planes really can turn into branes.

Compactifying further on K3, one finds a 4-dimensional theory that is
dual to F-theory on K3 $\times$ K3. The 24 7-branes wrapped over K3
become 24 anti-3-branes, which give rise to $-24$ units of tadpole in
the 4-form potential because of the WZ term. This is cancelled by
condensing 24 fundamental 3-branes in the 4d vacuum, as predicted in
\refs{\svw}. So we apparently have 24 anti-3-branes and 24 3-branes,
though the anti-branes (which are ``embedded'' in the 7-branes) arise
from the WZ coupling and do not break extra supersymmetry.

As is clear from the above discussion, this trend continues as we
compactify type I string theory down to lower dimensions. The Bianchi
identities for seven, six and five dimensional compactifications are
\eqn\lodibi{\eqalign{
d{^*G^{(6)}} &= {1\over 16\pi^2}~(\frac{1}{8}\tr R\wedge R - 
\tr F\wedge F)\delta(x^7)\delta(x^8)\delta(x^9)\cr
d{^*G^{(7)}} &= {1\over 16\pi^2}~(\frac{1}{16}\tr R\wedge R - 
\tr F\wedge F)\delta(x^6)\delta(x^7)\delta(x^8)\delta(x^9)\cr
d{^*G^{(8)}} &= {1\over 16\pi^2}~(\frac{1}{32}\tr R\wedge R - 
\tr F\wedge F)\delta(x^5)\delta(x^6)\delta(x^7)\delta(x^8)\delta(x^9).}}
We will have more to say about these Bianchi identities and their
relation to fractional charges in the next section.

The 7-dimensional case also shows some interesting features. In this
case there are 8 orientifold 6-planes, along with 16 6-branes. The
planes carry -2 units of magnetic charge with respect to the RR
1-form. The WZ term involving the RR 3-form $C^{(3)}$ is shared
equally between the branes and planes in this case. This suggests that
there may be a situation in which the orientifold planes can behave as
(single) branes, and indeed there is. Compactify the 7-dimensional
theory on K3 to 3 spacetime dimensions. This is dual to M-theory on K3
$\times$ K3, for which again it is known\refs{\svw} that the vacuum
contains 24 condensed 2-branes. These are precisely there to cancel
the 2-branes sitting ``inside'' the 16 6-branes and the 8 6-planes.

It is intriguing that apparently static objects like orientifold
planes actually contain so much dynamics.

\newsec{M-Theory on $S^1/Z_2$ and type IIB on $T^2/Z_2$}

In the previous section we saw how the Bianchi identity is modified in
the neighbourhood of an orientifold plane. It follows from these
modified Bianchi identities that the charges associated with RR
$p$-form potentials can be fractional. Orientifold planes are charged
with respect to these $p$-form RR potentials, and the charge
fractionalization due to modified Bianchi identities is closely
related to the fractional charges on the orientifold planes. In this
section we will show how these fractional charges are consistent with
the Dirac quantization condition in string theory and M-theory.

To illustrate this we will first consider M-theory compactified on
$S^1/Z_2$\refs{\hw}.  The $Z_2$ action here is an orientifold action
which takes $C \to -C$ and leaves other fields invariant. This theory
has to satisfy ten dimensional anomaly cancellation conditions at the
orientifold points of $S^1/Z_2$. The anomaly can be canceled by
putting 8 space filling nine branes at each end of the world. This
gives rise to the $E_8\times E_8$ heterotic string theory with each
end of the world contributing one $E_8$ gauge symmetry. In the M-theory
picture, these branes have no moduli and therefore they are stuck at
the two ends -- indeed, they are more like static planes than
dynamical branes, although one of our conclusions has been that there
is not so much of a distinction between the two objects.

We have already used the well-known result that the three form field
strength $H$ in the heterotic string theory satisfies the modified
Bianchi identity
\eqn\hbianchi{dH={1\over 16\pi^2}~( \tr R\wedge R- \tr F\wedge F)}
where $F$ is the gauge field strength taking values in $E_8\times E_8$
gauge group. This equation takes quite a different form in
M-theory. If we are close to one of the orientifold points on the
circle, only one of the two $E_8$ gauge symmetries is visible. At the
same time only half of the Pontryagin class of the curvature
contributes. Thus from this point of view, the Bianchi identity is
\eqn\halfbi{dH= {1\over 16\pi^2}~(\frac{1}{2}\tr R\wedge R- 
\tr F_1\wedge F_1),}
where the subscript 1 stands for one of the $E_8$ groups. When we lift
this equation to M-theory, the three-form field strength $H$ goes over
to the four-form field strength $G^{(4)}$ to give
\eqn\embi{dG^{(4)}={1\over 16\pi^2}~ (\frac{1}{2}\tr R\wedge R- 
\tr F_1\wedge F_1)\delta(x^{10}),}
where $x^{10}=0$ is the location of the orientifold
plane. Witten\refs{\witten} observed that since $R\wedge R$ is
quantized in integers, the magnetic charge of the four-form field
strength $G^{(4)}$ is quantized in half-integers.

The existence of such half-integral charge and its consistency with
Dirac quantization can be established by studying the world-volume
theory of a membrane\refs{\witten}. To do this let us first wrap the
world-volume of the membrane on a closed three cycle $T$ of the eleven
(or ten) dimensional manifold. What we want to find out is what
happens to the membrane path integral when we take it around a
circle. The WZ coupling of the membrane world volume theory to
$C^{(3)}$ is given by
\eqn\wzc{\exp(i\int_T C^{(3)})}
which when transported along the circle gives
\eqn\wzg{\exp(i\int_{T\times S^1} G^{(4)}).}
There is another factor which contributes to the phase of the membrane
path integral. This is related to the parity anomaly in the path
integral over world-volume fermions, from which it follows that
the interaction $\int C^{(3)}$ on the brane world-volume is modified
as
\eqn\modif{
C^{(3)} \rightarrow C^{(3)} + \half\tr(\omega_{3,L} + \omega_{3,N}) }
where $\omega_{3,L}$ and $\omega_{3,N}$ are the Chern-Simons 3-forms
associated respectively to the tangent and normal bundles to the brane
world-volume.

Thus the extra phase on transporting the membrane world-volume over a
circle is proportional to the first Pontryagin class of the normal
bundle to $T\times S^1$ (since the tangent bundle is trivial):
\eqn\extphase{
\exp(2\pi i [\half\int {\tr F\wedge F\over 16\pi^2}]) = 
\exp(i\pi {p_1(N)\over 2}) }
where $d\omega_{3,N} = {1\over 16\pi^2}~\tr F\wedge F = {p_1(N)\over
2}$.
 
Thus the total phase, which must be equal to
1, is given by\refs{\witten}
\eqn\mbphase{(-1)^{\int_{T\times S^1} {p_1(N)\over2}}
\exp(i\int_{T\times S^1} G^{(4)}).}
It follows that $G^{(4)}/2\pi$ has a half-integral period precisely on
those cycles on which the integral of $p_1(N)/2$ is odd. Therefore,
what is really observable is not the periods of $G^{(4)}/2\pi$ which
are half-integral but $G^{(4)}/2\pi -p_1(N)/4$ which is always
integral. In general, what appears in this condition is the full
Pontryagin class $p_1$ of the tangent bundle of the ambient
spacetime. Thus we see that the Dirac quantization condition is obeyed
by $G^{(4)}/2\pi -p_1/4$ charges.

As mentioned in the previous section, orientifold planes in the
$T^5/Z_2$ compactification of M-theory or of type IIA string theory
have half-integral magnetic charge with respect to $G^{(4)}$. If we
can find a four cycle in $T^5/Z_2$ where ${\tr(R\wedge R)\over
16\pi^2}= {p_1\over 2}$ integrates to an odd integer, this would give
us half integral charge. (In this case it is the Pontryagin class of
the tangent, rather than normal, bundle to the 4-manifold that
contributes.) The four cycle which encloses an orientifold fixed point
in $T^5/Z_2$ has this desired property. Though this cycle, $S^4/Z_2$,
is non-orientable, its Stiefel-Whitney class $w_4$, which is
equivalent to $R\wedge R$ mod 2 in the orientable case, is
unity\refs{\witten}.

Consider now the $T^6/Z_2$ compactification of type IIB string
theory. The modified Bianchi identity \bate\ in type IIB compactified
on $T^2/Z_2$ orientifold tells us that self-dual five-form charges are
quantized as $n-1/4$, where $n$ is an integer. The solution of the
Bianchi identity Eq.\bate\ relevant for this case can be written
\eqn\bianchsoln{
G^{(5)} = dC^{(4)} + \left( {1\over 4}\omega_{3,L} -
\omega_{3,Y}\right) \delta(x^8) \delta(x^9) }
or, alternatively, as
\eqn\altsoln{
G^{(5)} = dC^{(4)} +{1\over 16\pi^2}~ \left( {1\over 4}
\tr R\wedge R - \tr F
\wedge F \right) \times
\half \left( \epsilon(x^8) 
\delta(x^9) - \epsilon(x^9)\delta(x^8)\right) }
where $\epsilon(x)$ is the step function.

In contrast to the case of the $S^1/Z_2$ Bianchi, in this case neither
of the solutions is free of a $\delta$-function. This is related to the
fact that in the present case, spacetime does not just acquire an
end-of-the-world boundary but rather ends on submanifolds of
(real) codimension 2. 

The two solutions above differ by the addition to $G^{(5)}$ of the
exact form
\eqn\exactform{
d\left(\left( {1\over 4}\omega_{3,L} - \omega_{3,Y}\right) 
\times \half \left( \epsilon(x^8) \delta(x^9) - \epsilon(x^9)
\delta(x^8)\right)\right)}

The $T^6/Z_2$ orientifold compactification is the place where we
expect that the charge carried by the orientifold planes is $-1/4$. To
measure this charge, we consider a five cycle, $S^5/Z_2$ which
encloses the orientifold fixed point and integrate the self-dual
five-form $dC^{(4)}$ over this five cycle. As is clear from Eq.\bate,
in the absence of gauge fields there is another contribution coming
from the term ${1\over 4}{\tr(R\wedge R)\over 16\pi^2}\delta(x^5)$.
Since we are only interested in fractional charges we can safely
ignore the gauge field contribution which is always an integer. The
charge of the orientifold plane, which is $-1/4$, is obtained by
integrating the five-form $dC^{(4)}$ over $S^5/Z_2$. What remains to
be done, by analogy with similar manipulations in Ref.\witten, is to
integrate ${1\over 16\pi^2}~\tr(R\wedge R)\delta(x^5)/4$ over
$S^5/Z_2$.

At this point we give a more general argument which 
explains the existence of fractional charge when we
integrate the quantity
\eqn\tin{T\equiv \frac{1}{2^m}~{\tr(R\wedge
R)\over 16\pi^2}~\delta(x_i)\delta(x_j)...\delta(x_p).}
over the manifold $S^{4+q}/Z_2$. Here $m, i, j, ... ,p$ are the
relevant integers and $q$ is the number of delta functions. For the
$T^6/Z_2$ case $m = 2, i = 5$ and $q = 1$.
 
$S^{4+q}$ is defined by
\eqn\sph{S^{4+q}: x_1^2 + x_2^2 + ... + x_{4+q}^2 + x_{5+q}^2 = 1.}
And $S^{4+q}/Z_2$ is defined modding out with the antipodal map $x_i
\to -x_i$ for all $i$. Now the integration is simple to perform. The
$\delta(x_i), \delta(x_j), ...$ factors fix us at the locus $x_i = 0,
x_j = 0, ...$ This locus is a section of $S^{4+q}/Z_2$ which is
nothing but $S^4/Z_2$ because there are $q$ delta functions. It is
well known that $S^4/Z_2$ can be naturally embedded in
$S^{4+q}/Z_2$. This embedding corresponds to setting $q$ coordinates
of $S^{4+q}/Z_2$ to zero. Integrating out the delta function precisely
implements this action\footnote{$^2$} {The simplest geometrical way to
think of it is that the boundary of a ball in 3 dimensions is $S^2$,
and its equator is $S^1$. A $Z_2$ modding will make the equator
$S^1/Z_2$. So if there is a quantity to be integrated over $S^2/Z_2$
with a delta-function along say the $z$ direction, it will reduce to
an integral over $S^1/Z_2$. This also seems like $S^2/Z_2$ being
represented as a ``fibration'' over $S^1/Z_2$ with a fibre
$S^1$. Since the $Z_2$ action has done nothing to the fibre the
integral of delta function will be just 1.}.  Therefore now we only
have to integrate ${1\over 2^m}~{\tr(R\wedge R)\over 16\pi^2}$ over
$S^4/Z_2$.

The quantity $\lambda \equiv  {1\over 16\pi ^2}~\tr(R\wedge R)$ is
congruent modulo two to the Stiefel-Whitney class $w_4$. By a
standard computation\refs{\mill} one can show that
\eqn\stas{\int_{S^4/Z_2}~~w_4 =~ 1~~~ mod~~ 2.}
Together with the ${1\over 2^m}$ factor, this would point to the
existence of fractional fluxes for the corresponding fields. For the
$T^6/Z_2$ example considered we see that the period of the five-form
$G^{(5)}$ is fractional precisely on those five-cycles on which the
integral of ${1\over 4}~{\tr(R\wedge R)\over 16\pi^2}\delta (x^5)$
contributes the compensating fraction, so that the total charge is
effectively integer and the Dirac quantization condition is then
satisfied by the charges of the field $G^{(5)} - {1\over
4}~{\tr(R\wedge R)\over 16\pi^2} \delta(x)$.\footnote {$^3$} {In case
of $S^5/Z_2$, there is another way to show how $G^{(5)}-{1\over
4}~{\tr(R\wedge R)\over 16\pi^2}\delta(x)$ satisfies the Dirac
quantisation condition. To see this we use the fact that $S^5$ is a
generalized Hopf fibration over $CP^2$ with $S^1$ fibre. The antipodal
map which takes $S^5$ to $S^5/Z_2$ acts trivially on $CP^2$ but it
halves the volume of the fibre. Therefore, $S^5/Z_2$ is also an $S^1$
fibration over $CP^2$. The difference between these two bundles is
that the Chern class of the latter is double of that of the former.
To evaluate ${1\over 4}~{\tr(R\wedge R)\over 16\pi^2}\delta(x^5)$, one
first integrates along the fibre to reduce the top form on $S^5/Z_2$
to the top form on the base, i.e., $CP^2$.  Integration along the
fibre can be done if the cohomology classes have compact support in
the vertical direction\refs{\bottu}. Since the Chern class is doubled,
integration along the fibre gives $${1\over 4}~{\tr(R\wedge R)\over
16\pi^2}|_{CP^2}\int_{S^1}\delta(x^5)dx^5={1\over 2}~ {\tr(R\wedge
R)\over 16\pi^2}|_{CP^2}.$$ The RHS of the equation can be written in
terms of the Pontryagin class of $CP^2$ as $p_1(R)/4$. Thus the
integration along the fibre gives the top form on $CP^2$ which is one
quarter of its first Pontryagin class. Since the integral of $p_1(R)$
over $CP^2$ is equal to 3, integrating it on $CP^2$ one obtains the
contribution of the curvature terms, which is equal to $3/4$.}  The
crucial thing that has entered into the discussion is that the object
which we want to integrate has support only a $S^4/Z_2$ submanifold of
$S^{4+q}/Z_2$.

It is interesting to consider whether other physical effects are
related to the occurrence of the fractional fluxes that we have been
discussing. For example, fractional fluxes similar to those discussed
above would appear in other orientifolds based on higher discrete
groups than $Z_2$ and correspondingly with lower
supersymmetry. However, there do seem to be some important
distinctions between the situation discussed in the context of
$S^1/Z_2$ orientifolds\refs{\hw} and the more general cases considered
here. In the former case, one can write the modified Bianchi identity
at an orientifold plane as a completely nonsingular boundary
condition, while in the latter cases it needs to be expressed in a
singular form. Presumably related to this is the fact that the former
case has wider applicability: besides orientifold planes, fractional
flux for a 4-form field strength can occur in M-theory
compactifications on any complex 4-fold with a 4-cycle whose first
Pontryagin class is a multiple of 2 but not 4. It remains to be seen
whether charges of the type ${1\over 2^n}$ for field strengths
$G^{(n+3)}$ can be realized in smooth compactifications, but from the
present considerations this does not seem likely. In the same vein,
$p$-branes for $p\ge 3$ do not have anomalies with the right discrete
ambiguity $Z_{2^{p-1}}$ to play a role similar to the parity anomaly
on the 2-brane. It remains to understand what exactly happens to the
parity anomaly on the 2-brane when it is T-dualized to a higher
brane. Comments in this direction appear in
Refs.\refs{\seitwo}\refs{\seifour}\refs{\seiwitthree} but a careful
analysis remains to be carried out.

In this section we have seen how the fractional charges on the
orientifold planes, which could potentially be inconsistent with the
Dirac quantization condition, actually conspire to give consistent
results in the case of $T^n/Z_2$ orientifolds with $n\ge 5$. In
subsequent sections we examine other aspects of gravitational
couplings and some details of low-dimensional orientifolds.

\newsec{$R^4$ Wess-Zumino terms on high-dimensional branes and planes}

We have seen that $R^2$ couplings of Wess-Zumino type appear on
orientifold planes as well as branes. Here we will show that the same
is true for $R^4$ couplings, though for obvious reasons these can only
appear on $p$-branes and $p$-planes for $p\ge 7$.

In 10 dimensions, the type I string has a term $B\wedge X_8$ which
plays an essential role in the Green-Schwarz anomaly cancellation
mechanism. Here $X_8$ is the 8-form
\eqn\xeight{
\eqalign{
X_8 &= {1\over (2\pi)^4}\left({1\over 48}\tr F^4-{1\over 192}\tr F^2\tr R^2 
+ {1\over 384}\tr R^4 + {1\over 1536} (\tr R^2)^2\right)\cr
&= {1\over (2\pi)^4}\left({1\over 48}\tr F^4-{1\over 192}\tr F^2\tr R^2 \right)+
\tilde X_8 },}
where,
\eqn\txeight{\tilde X_8 ={1\over 128} (p_1)^2 - {1\over 96} p_2 .}
The first Pontryagin class $p_1$ was defined earlier, and the second
Pontryagin class is given by
\eqn\ptwo{
p_2 = {1\over (2\pi)^4}\left(-{1\over 4} \tr R^4 + {1\over 8}(\tr
R^2)^2\right) }  
Let us first look at the terms which contain gauge fields. It is easy to see 
that these terms can be obtained by expanding the Wess-Zumino terms on the 
D-branes. In case of multiple coincident branes all we need to do is to define 
the trace in the fundamental representation of the appropriate gauge group 
generated by coincident branes. In the case at hand, we have 16 coincident 
D9-branes in the presence of an O9-plane, which leads to SO(32) gauge symmetry.
The terms involving gauge fields obtained from expanding the Wess-Zumino term 
on the world volume correctly reproduce the $F^4$ and $F^2R^2$ terms in $X_8$.
Hence, as one would expect, there is no need to assign any gauge couplings
to the orientifold plane.

Now we will address the analogous issue for the $R^4$ terms. This
time it will prove necessary to assign specific couplings to the
09-plane, as was the case for $R^2$ terms.
As before, we decompose this term into the contribution from the bulk,
the branes and the orientifold plane, all of which are of course
coincident in 10 dimensions. No term of the above form is present in
type IIB in 10 dimensions. The contribution on a single 9-brane, which
we denote $B\wedge B_8$, is extracted from the anomaly inflow formula, which
leads to
\eqn\beight{
B_8 = {1\over 320}\left({1\over 8}(p_1)^2 - {1\over 9} p_2 \right) }
This can be conveniently recast in terms of $\tilde X_8$ and $\tr R^4$, and
we find
%
%
%
\eqn\beightre{
B_8 = {1\over 16}\left( {4\over 3}\tilde X_8 - {1\over 480} {\tr R^4\over
(2\pi)^4} \right) }
Since we require that the contribution from 16 branes plus that from
the plane must provide the total $R^4$ term, we have $16 B_8 + P_8 =
\tilde X_8$, where $B\wedge P_8$ is the plane contribution. $P_8$ is found to
be\footnote{$^4$} {Note that the following equation differs from the
expression that appeared in the original version and the published
form of this paper. This is due to an error of a factor of 2 in the
normalisation of Eq.\xeight\ above. We were motivated to check and
revise these expressions due to the appearance of some recent
preprints\refs{\morales,\stefanski}\ (see also \refs\craps), in which
our proposal for $R^4$ couplings on orientifold planes was checked by
explicit computation of string amplitudes. The computations confirmed
that such couplings exist on O-planes as we had predicted, but showed
an error in our precise coefficients. The following equation,
Eq.(4.6), now agrees with the conclusions of these papers.}
\eqn\peight{
P_8 = -{1\over 3}\tilde X_8 + {1\over 480} {\tr R^4\over (2\pi)^4}. }
This can be re-expressed in terms of Pontryagin classes as
\eqn\peightre{
P_8 = {1\over 640}(p_1)^2 - {7\over 1440}p_2 }

One might think that Eqs.\beightre\ and \peight\ together contradict the
fact that a 7-plane can split into a pair of 7-branes. However, the
$R^4$ terms on 7-branes and 7-planes are of the form
${\tilde\phi}\wedge B_8$ and ${\tilde\phi}\wedge P_8$ respectively,
where $\tilde\phi$ is the RR scalar. When 7-planes split into $(p,q)$
7-branes, since $\tilde\phi$ is not SL(2, Z) invariant one cannot say
what the $(p,q)$ branes should carry. In this respect the situation is
similar to that for the 8-form charge carried by 7-planes and
7-branes, which according to Ref.\sen\ does not split additively
because of the non-Abelian nature of the monodromy. This is in
contrast to the $R^2$ term, where the RR potential $C^{(4)}$ that
appears is SL(2, Z) invariant, and the term splits additively.

\newsec{Gravitational couplings in $3+1$ dimensional gauge theory}

It has been observed that certain supersymmetric gauge theories in
$3+1$ dimensions have partition functions which are modular under
$SL(2,Z)$ with nontrivial weight. The resolution to this apparent
failure of exact $SL(2,Z)$ invariance, or modular anomaly, is that
these theories have specific couplings to gravity which produce
(cancelling) modular anomalies.

Here we will realize the relevant gauge theories on world-volumes of
3-branes, and will investigate the relationship between the
gravitational couplings required for consistency of gauge theories and
those generated by branes and planes on their world-volumes.

Consider $N=4$ super-Yang-Mills in 3+1d. This is the world-volume
field theory of a 3-brane. It can be considered to be topologically
twisted (namely, the physical and twisted theories are equivalent),
when written on flat spacetime or (after Euclideanization) on
hyper-K\"ahler 4-manifolds. As we have seen, the gravitational
Wess-Zumino couplings on the 3-brane world-volume are known to be
\eqn\gravthree{
{1\over 16\pi^2}~{1\over 24}\tr\left( C^{(0)} R \wedge R \right) }
where $C^{(0)}$ is the RR scalar of type IIB. However, this coupling
is not $SL(2,Z)$ invariant or even covariant. 

To discover the correct extension of the above coupling, we need to
realize a known vacuum of string theory in terms of condensed
branes. The appropriate vacuum in this case is the orientifold of type
IIB on $T^6/Z_2$, which has already made an appearance above. This
vacuum has $N=4$ spacetime supersymmetry, and it gauge sector is an
$N=4$ super-Yang-Mills theory. In this way of describing the vacuum,
there are 16 3-branes along with 64 orientifold planes. As far as
world-volume gravitational couplings are concerned, we have argued
that the planes carry $1/2$ the fraction carried by the branes, so
that to find the contribution on a single brane world-volume, we need
to divide relevant terms in the spacetime action by 24.

In Ref.\refs{\hm} it is observed that the spacetime $R^2$ coupling
in 4d $N=4$ compactifications is, at tree level\footnote{$^5$} {Our
conventions differ slightly from those in Ref.\refs{\hm}, since we
want to make the anomaly term purely holomorphic rather than
anti-holomorphic.}, proportional to

\eqn\rsquare{
\tr \left( C^{(0)} R \wedge R + e^{-\phi} R \wedge {^*R} \right)}
This is argued in Ref.\refs{\hm} to be corrected by the replacement
\eqn\corrected{
\eqalign{
C^{(0)}&\rightarrow {\rm Re}({\log \eta(\tau)^{24}\over 2\pi i})\cr
e^{-\phi} &\rightarrow {\rm Im}({\log \eta(\tau)^{24}\over 2\pi i})\cr}}
where $\tau\equiv C^{(0)} + i e^{-\phi}$. In the limit of constant
dilaton and axion, the action gets a contribution depending only on
the topological invariants $\chi$ (the Euler characteristic) and
$\sigma$ (the signature) of the spacetime 4-manifold:
\eqn\actiontopinv{
-(\chi - {3\over2}\sigma)\log \eta^{12}
-(\chi + {3\over2}\sigma)\log {\bar\eta}^{12}}

It follows that each brane carries ${1\over 24}$ of this term. To
leading order and considering only the term involving $\sigma$, this
precisely coincides with Eq.\gravthree, given that $\sigma=p_1/3$
where $p_1$ is the correctly normalized first Pontryagin class. The
full gravitational coupling on the brane, to second order in
derivatives, is thus
\eqn\fullgrav{
-{1\over 4}(2\chi - 3\sigma)\log\eta
-{1\over 4}(2\chi + 3\sigma)\log{\bar \eta}}
One way to check that this is correct is to note than on
hyper-K\"ahler manifolds, the modular anomaly from this must cancel
that coming from the gauge partition function, for which we have the
result\refs{\vawitt}:
\eqn\gaugeanom{
Z_{gauge}({-{1\over\tau}}) = \tau^{\chi\over2}Z_{gauge}(\tau)}
To check cancellation of the modular anomaly, we examine the $SL(2,Z)$
transformation law for Eqn.\fullgrav\ after setting
$\sigma=-{2\over3}\chi$ which is the case for hyper-K\"ahler
manifolds. Thus we need to know how the term in the functional
integral
\eqn\functterm{
Z_{grav} = \exp(-\chi\log\eta)= \eta^{-\chi} }
transforms. Using $\eta(-{1\over\tau}) = \tau^{1\over 2}\eta(\tau)$ we
find that the gravitational contribution to the modular anomaly is
\eqn\gravanom{ 
Z_{gauge}({-{1\over\tau}}) = \tau^{-{\chi\over2}}Z_{gauge}(\tau) }
which exactly cancels the gauge contribution.

\newsec{Some Issues Concerning Fractional Charge in Dimensions $d<3$}

Although not specifically related to gravitational couplings on branes
and planes, there is a curious situation in which gravitational
anomalies turn into fractional charges on planes upon
compactification. We discuss this below and explain how chiral
supersymmetry in this problem cures an apparent paradox.

Consider the orientifold of type IIA on $T^9/Z_2$. By T-duality, this
vacuum, where all of space is compactified, is realized with 16
0-branes located at points in the internal torus (of course in such
low dimensions the concept of moduli space is not strictly
appropriate). The $2^9 = 512$ orientifold points each carry a charge
$-{1\over 32}$ with respect to the RR 1-form. This vacuum may be
considered a limit of M-theory on $T^9/Z_2$ to 2 spacetime dimensions,
further compactified on a circle, as the circle shrinks. However, in
the M-theory case\refs{\kdsmorb,\senD,\kdsmI} one expects 512 chiral
fermions to appear in the twisted sector, located symmetrically at the
512 fixed points.

Thus it would appear that on compactification of the M-theory
orientifold on a further circle, 512 chiral fermions in $1+1$
noncompact dimensions must suddenly turn into 16 D0-branes, while the
gravitational anomaly carried by each of the $2^9$ fixed planes (which
are really fixed lines) turns into $-{1\over 32}$ units of 1-form
charge. The fermions in $1+1$ dimensions were forced to sit at the
fixed points to bring about local gravitational-anomaly cancellation,
so it is hard to understand how they go over into 16 objects in $0+1$
dimensions which apparently cannot bring about local 1-form charge
cancellation.

The resolution to this lies in the supersymmetry of this
problem. In $1+1$ dimensions, the above orientifold of M-theory has
$(0,16)$ {\it chiral} supersymmetry. The algebra is:
\eqn\chiralsuper{
\{Q_-^i, Q_-^j \} = \delta_{ij} P_- }
The chiral fermions which appear as twisted sectors are singlets of
supersymmetry, which means they have $+$ chirality in these
conventions, and hence are annihilated by both sides of the
algebra by virtue of the Dirac equation $P_- \psi_+ = 0$.

On compactification to $0+1$ dimensions, we end up with a
supersymmetric quantum mechanics that is also chiral, in the sense
that now the supercharges $Q^i$ satisfy
\eqn\chiraloned{
\{Q^i, Q^j \} = \delta_{ij} (P-Z) }
where $Z$ is a central charge. D 0-branes are BPS, which means they
are annihilated by the RHS of this algebra. Thus in fact the 0-branes
propagating in the fully compactified space are singlets of the
residual supersymmetry, rather than multiplets with 32 states (16
bosonic, 16 fermionic) as they are in higher dimensions. As a result,
in type IIA on $T^9/Z_2$ the twisted sector of the orientifold can be
thought of as being made up of 512 supersymmetry singlets, and local
gravitational anomaly cancellation goes directly over into local
charge cancellation.

\newsec{Discussion}

We have seen that curved orientifold planes carry gravitational
couplings of WZ type. Quite plausibly they also carry other types of
gravitational couplings such as $R^4$ terms (not of Wess-Zumino type)
or their dimensional reductions. This would be interesting to
investigate, along with the possible appearance of similar terms on
curved D-branes.

It has also been argued here that fractional fluxes can consistently
be carried by orientifold planes. The general conclusion from this
discussion would be that such phenomena as Dirac quantization can be
modified by suitable (topological) gravitational couplings, as first
noted in Ref.\refs{\witten}.

While all this adds some insight into the fascinating interplay
between gauge and gravitational interactions in string and M-theory, a
deeper understanding of this interplay would be desirable. Also, it
would be interesting to understand gravitational couplings in $N=2$
supersymmetric gauge theories\refs{\witab} from the point of view of
3-branes in suitable backgrounds.
\bigskip

\noindent{\bf Acknowledgements:} We acknowledge helpful discussions 
with I. Biswas, A. Dabholkar, R. Dijkgraaf, M. Douglas, S.F. Hassan,
C. Imbimbo, K.S. Narain, N. Nitsure, N. Raghavendra, A. Sen,
W. Taylor, E. Verlinde, H. Verlinde and E. Witten. DJ would like to
thank the Tata Institute of Fundamental Research for hospitality.  SM
is grateful for support from the Nederlandse Organisatie Voor
Wetenschappelijk Onderzoek and the hospitality of Herman Verlinde and
the Universiteit van Amsterdam.

\listrefs

\bye